# Anomalous Ettingshausen effect in iron-carbon alloys

Ren Nagasawa[1,2,†], Koichi Oyanagi[1], Takamasa Hirai[2], Rajkumar Modak[2], Satoru Kobayashi[1], and Ken-ichi Uchida[2,3]

**AFFILIATIONS**

[1] Faculty of Science and Engineering, Iwate University, Morioka 020-8851, Japan.

[2] National Institute for Materials Science, Tsukuba 305-0047, Japan.

[3] Institute for Materials Research, Tohoku University, Sendai 980-8577, Japan.

Authors to whom correspondence should be addressed: k.0yanagi444@gmail.com, UCHIDA.Kenichi@nims.go.jp

[†] Present address: Graduate School of Pure and Applied Sciences, University of Tsukuba, Tsukuba 305-8573, Japan.

**ABSTRACT**

We have investigated the anomalous Ettingshausen effect (AEE) in iron-carbon alloys, i.e., cast irons and steel, using the lock-in thermography. All the alloys exhibit the clear AEE-induced temperature modulation, and their anomalous Ettingshausen coefficient is an order of magnitude greater than that of the pure iron at room temperature. The dimensionless figure of merit for AEE in the ductile cast iron is 55 times greater than that in the pure iron owing to the significant increase of the anomalous Ettingshausen coefficient. Our result reveals a potential of iron-carbon alloys as transverse thermoelectric materials, although the composition and microstructures optimizations are necessary.

Transverse magneto-thermoelectric effects[1-3] refer to the conversion of a heat current into a transverse charge current, or vice versa, in a magnetic material with spontaneous magnetization. The orthogonal relationship between the input and output currents enables the construction of thermoelectric devices with simple lateral structure, convenient scalability, and easy fabrication, which cannot be achieved using the conventional thermoelectric effects, i.e., the Seebeck and Peltier effects. One of the representative transverse magneto-thermoelectric effects is the anomalous Ettingshausen effect (AEE).[16-39] AEE refers to the conversion of a charge current into a transverse heat current in a magnetic material[4-15], which is the reciprocal of the anomalous Nernst effect (ANE). The AEE-induced heat current density $\mathbf{j}_{q,\text{AEE}}$ is described as

$$\mathbf{j}_{q,\text{AEE}} = \Pi_{\text{AEE}}(\mathbf{j}_c \times \mathbf{m}), \qquad (1)$$

where $\mathbf{j}_c$, $\mathbf{m}$, and $\Pi_{AEE}$ denote the charge current density, unit vector of magnetization, and anomalous Ettingshausen coefficient, respectively [Fig. 1(a)]. Since AEE works as a temperature modulator with simple structure and versatile scaling owing to the unique thermoelectric conversion symmetry, it may pave the way for thermal management technologies for electronic and spintronic devices. However, at present, AEE is far from practical applications because the thermoelectric conversion performance is still several orders of magnitude smaller than that of the conventional Peltier effect. Therefore, the materials development to get large AEE is important for thermoelectric applications based on AEE.

Recently, large AEE and ANE were observed in various iron-based compounds and alloys.[16-18] Since iron is the highest abundance among the magnetic elements and nontoxic element, iron-based materials are promising, while pure iron exhibits small AEE/ANE.[5,16-19] Up to now, AEE and ANE have been investigated in FeAl,[16,17] FePt,[20,21,24,36] FeNi,[22,28] $Fe_3O_4$,[23] FePd,[24,36] FeGa,[16,18,27] $Fe_4N$,[29] $Fe_3Sn_2$,[35] FeSi,[37] FeRh,[38] and FeGd.[39] Interestingly, these results show that the amplitude and sign of the effects in the iron-based materials sensitively depend on the alloy composition and elements. However, the transverse magneto-thermoelectric properties of the iron-carbon alloys have not been investigated. Carbon is the typical element added to iron to improve its hardness,[40,41] and the iron-carbon alloys, such as cast irons and steel, are widely used in industry.

In this study, we report the observation of AEE in iron-carbon alloys at room temperature by means of the lock-in thermography (LIT) method.[7-15,42-46] We found that commercially available dense bars of the iron-carbon alloys show clear AEE-induced temperature modulation. The maximum absolute value of AEE in cast irons is an order of magnitude greater than that of pure irons. We also measured the electrical and thermal conductivities and found that the dimensionless figure of merit for AEE/ANE in a cast iron is 55 times larger than that in pure iron.

We commercially obtained three different types of iron-carbon alloys: the ductile cast iron and gray cast iron from Kogi Corp., and steel from Kobe Steel, Ltd. Our ductile and gray cast irons contain a carbon content of > 13 at.%, while the steel contains that of < 1 at.% (see the detailed compositions of the iron-carbon alloys in Table I). However, because of the solubility limit of the iron-carbon binary equilibrium phase diagram at room temperature, the actual carbon concentration of the main phase in the cast irons is about 8 at.%, and the excess carbon forms graphite deposits in the matrix. The ductile cast iron, gray cast iron, and steel show the tensile strength larger than 400, 330, and 400 $Nmm^{-2}$, respectively. For the measurements of the electrical conductivity $\sigma$, thermal diffusivity $\alpha$, and AEE, we cut the iron-carbon alloy bars into rectangular with the dimension of 1.5 × 15.0 × 1.0, 1.5 × 10.0 × 10.0, and 1.5 × 15.0 × 1.0 mm, respectively, with an electrical discharge machining system. We measured $\sigma$ by the four-probe method and determined the thermal conductivity $\kappa$ using $\alpha$ measured by the laser flash method, specific heat measured by the differential scanning calorimetry, and density measured by the Archimedes method. The magnetization $M$ of the alloys was measured by vibrating sample magnetometry.

We investigated AEE in the iron-carbon alloys using the LIT method,[7-15,42-46] which allows us to estimate $\Pi_{AEE}$ [Fig. 1(b)]. We measured thermal images of the surface of the iron-carbon alloys using an infrared camera with applying a square-wave-modulated AC charge current with the frequency $f$, amplitude $J_c$ (with the density $j_c$), and zero offset to the sample along the $y$ direction and an in-plane magnetic field $H$ along the $z$ direction. We obtained the first harmonic response of the detected images, which are transformed into the lock-in amplitude $A$ and phase $\phi$ images through Fourier analysis. Based on this procedure, we obtain thermoelectric signals ($\propto J_c$) free from Joule-heating signals ($\propto J_c^2$). The AEE signals can be separated from parasitic signals due to the Peltier effect generated at the ends of the sample by measuring the $H$ dependence of the LIT images because the sign of the AEE-induced heat current is reversed by reversing **m** [Eq. (1)]. To enhance the infrared emissivity, the top surface of the sample was coated with insulating black ink. We carried out all the measurements at room temperature and atmospheric pressure.

Figures 1(c) and 1(d) show the $A$ and $\phi$ images for the ductile cast iron at the temperature $T$ = 301 K measured with applying the charge current of $f$ = 5.0 Hz and $J_c$ = 1.0 A along the $y$ direction, where $T$ was estimated from steady-state thermal images. Clear current-induced temperature modulation appears on the entire surface of the sample with $\mu_0 H$ = +1.8 T [see the left images of Figs. 1(c) and 1(d)]. The signal with $\phi \sim 0°$ means that the top surface of the sample is heated when the charge current is along the +$y$ direction. With the $H$ reversal (at $\mu_0 H$ = −1.8 T), although the $A$ distribution does not change, the phase reversal appears, where $\phi \sim$ 180° [see the right images of Figs. 1(c), and 1(d)], indicating that the sign of the current-induced temperature modulation is reversed by reversing the magnetic field. This behavior is consistent with the symmetry of AEE [see Eq. (1)]. We carried out the same experiments also for the gray cast iron and steel, and obtained qualitatively same results, indicating the appearance of AEE in these iron-carbon alloys. To extract the pure AEE contribution, hereafter we focus on the $H$-odd-dependent component of the current-induced temperature modulation, where the LIT signals are discussed in terms of $A_{\text{odd}} = |A(+H)\exp\{-i\phi(+H)\} - A(-H)\exp\{-i\phi(-H)\}|/2$ and $\phi_{\text{odd}} = -\arg[A(+H)\exp\{-i\phi(+H)\} - A(-H)\exp\{-i\phi(-H)\}]$,[46] and show only the $A_{\text{odd}}$ values because we confirmed that $\phi_{\text{odd}}$ is always close to 0° in the similar manner to the left image in Fig. 1(d).[13,15,45]

To check the behaviors of AEE, we measured $A_{\text{odd}}$ with changing the magnitude of the applied charge current and magnetic field. Figure 2(a) shows the $J_c$ dependence of the $A_{\text{odd}}$ signal for the ductile cast iron, gray cast iron, and steel at $f = 5.0$ Hz and $\mu_0|H| = 1.8$ T. In all the samples, the magnitude of the $A_{\text{odd}}$ signal is proportional to $J_c$. Figure 2(b) shows the $A_{\text{odd}}$ signal as a function of $\mu_0|H|$ at $f = 5.0$ Hz and $J_c = 1.0$ A. The magnitude of the $A_{\text{odd}}$ signal increases with increasing $\mu_0|H|$ and saturates at ~1.0 T; the $\mu_0|H|$ dependence of $A_{\text{odd}}$ follows the magnetization curve of the sample [solid curves in Fig. 2(b)]. These results are consistent with the expected behavior of AEE. We thus conclude that all the iron-carbon alloys exhibit AEE.

For the quantitative discussion, we estimated the magnitude of the AEE-induced temperature modulation in the steady state and determined $\Pi_{\text{AEE}}$. Figure 2(c) shows the $f$ dependence of the $A_{\text{odd}}$ signals for the ductile cast iron, gray cast iron, and steel at $J_c = 1.0$ A and $\mu_0|H| = 1.8$ T. The magnitude of the $A_{\text{odd}}$ signal gradually decreases with increasing $f$ due to the suppression of thermal diffusion.[11,42,45] In contrast, the magnitude of the temperature modulation signals keeps a constant value in the low $f$ region [see the gray shading in Fig. 2(c)], indicating that the AEE-induced temperature modulation reaches the steady state for $f \leq 1.0$ Hz. We determined the AEE-induced temperature modulation at the steady state $A_{\text{odd}}^{\text{steady}}$ by averaging the $A_{\text{odd}}$ signals at $f = 1.0, 0.5,$ and $0.2$ Hz. From the steady-state AEE signal, we can determine the anomalous Ettingshausen coefficient[14] as

$$\Pi_{\text{AEE}} = \frac{\pi}{4}\frac{\kappa \Delta T_{\text{AEE}}}{j_c L}, \qquad (2)$$

where $\Delta T_{\text{AEE}}$ and $L$ are the AEE-induced temperature difference between the top and bottom surfaces of the sample, i.e., $|\Delta T_{\text{AEE}}| = 2A_{\text{odd}}^{\text{steady}}$, and thickness of the sample (= 1.5 mm), respectively. Here, the sample is assumed to be thermally isolated, and the sign of $\Delta T\_\text{AEE}$ is determined by the phase value. We can also estimate the anomalous Nernst coefficient $S_{\text{ANE}}$ via the Onsager reciprocal relation[8]: $\Pi_{\text{AEE}} = S_{\text{ANE}}T$.

In Fig. 3(a), we summarize $\Pi_{\text{AEE}}$ and corresponding $S_{\text{ANE}}$ of the ductile cast iron, gray cast iron, and steel estimated from the above-mentioned procedures, and of the pure iron from Ref. 13. The absolute value of $\Pi_{\text{AEE}}$ of the cast irons, $\Pi_{\text{AEE}} = 2.5 \times 10^{-4}$ V for the ductile cast iron and 2.3 × 10$^{-4}$ V for the gray cast iron, is an order of magnitude larger than that for the pure iron, $\Pi_{AEE}$ = –2.4 × 10$^{-5}$ V. The sign of $\Pi_{AEE}$ and $S_{ANE}$ of the cast irons is the positive, which is the same as that of other iron-based compounds and alloys,[16-18] while that of the steel and pure iron is negative. We plot $|\Pi_{AEE}|$ as a function of the saturation magnetization $M_s$ of the iron-carbon alloys and pure iron in the inset to Fig. 3(a). Despite the crucial difference in $|\Pi_{AEE}|$, all the alloys and pure iron show the almost same $M_s$; $\Pi_{AEE}$ has no correlation with $M_s$.

Figure 3(b) shows the dimensionless figure of merit for AEE/ANE in the iron-carbon alloys, which was estimated based on the following definition:[11,14,47]

$$Z_{AEE}T = \frac{\Pi_{AEE}^2 \sigma}{\kappa}\frac{1}{T}\left(=\frac{S_{ANE}^2 \sigma}{\kappa}T\right). \quad (3)$$

By using $\Pi_{AEE}$ [Fig. 3(a)] together with $\sigma$ [Fig. 3(c)] and $\kappa$ [Fig. 3(d)], we obtained the maximum value of $Z_{AEE}T$ = 1.3 × 10$^{-5}$ in the ductile cast iron, which is 55 times larger than that of the pure iron[13] ($Z_{AEE}T$ = 2.4 × 10$^{-7}$). The larger $Z_{AEE}T$ in the cast irons is attributed to the larger $\Pi_{AEE}$. Here, we note that even the maximum $Z_{AEE}T$ value for the iron-carbon alloys used in this study is more than an order of magnitude smaller than the record-high values at room temperature, obtained in SmCo$_5$-type permanent magnets[11] and full-Heusler Co$_2$MnGa.[30-33]

The AEE in the cast irons is determined mainly by the off-diagonal component of the thermoelectric conductivity tensor, i.e., the transverse thermoelectric conductivity. In general, the anomalous Ettingshausen coefficient is composed of two components[11]: $\Pi_{AEE}$ = ($\rho_{xx}\alpha_{xy}$ + $\rho_{xy}\alpha_{xx}$)$T$ = $\Pi_I$ + $\Pi_{II}$, where $\rho_{xx}$ ($\rho_{xy}$) is the diagonal (off-diagonal) component of the electric resistivity tensor, $\alpha_{xx}$ ($\alpha_{xy}$) the diagonal (off-diagonal) component of the thermoelectric conductivity tensor, $\Pi_I$ = $\rho_{xx}\alpha_{xy}T$, and $\Pi_{II}$ = $\rho_{xy}\alpha_{xx}T$. $\Pi_I$ and $\Pi_{II}$ are regarded as an intrinsic part of AEE and concerted action of the Seebeck and anomalous Hall effects, respectively. Figure 4 shows the $\Pi_I$ and $\Pi_{II}$ values at 301 K estimated for the cast irons; $\Pi_I$ and $\Pi_{II}$ are respectively estimated to be 2.5 × 10$^{-4}$ V and 0.1 × 10$^{-4}$ V (2.2 × 10$^{-4}$ V and 0.1 × 10$^{-4}$ V) for the ductile (gray) cast iron, indicating the dominant $\Pi_I$ contribution. With the experimental results of the longitudinal resistivity ($\rho_{xx}$ = $\sigma^{-1}$), Seebeck coefficient ($S_{xx}$ = $\rho_{xx}\alpha_{xx}$), and anomalous Hall angle ($\theta_{AHE}$ = $\rho_{xy}/\rho_{xx}$), we estimated the transverse thermoelectric conductivity $\alpha_{xy}$ in the ductile (gray) cast iron as 1.4 K$^{-1}$Am$^{-1}$ (1.2 K$^{-1}$Am$^{-1}$), where $\rho_{xx}$ = 5.7 × 10$^{-7}$ Ωm (6.3 × 10$^{-7}$ Ωm), $S_{xx}$ = -7.2 × 10$^{-6}$ VK$^{-1}$ (-6.5 × 10$^{-6}$ VK$^{-1}$), and $\theta_{AHE}$ = -3.7 × 10$^{-3}$ (-3.6 × 10$^{-3}$). Although the origin of $\alpha_{xy}$ in the cast irons is yet to be clarified, detailed calculations of electronic band structures will provide an insight into the microscopic mechanism of the enhancement of AEE in the cast irons.

The difference in the physical properties between the ductile and gray cast irons is due to the morphology of graphite precipitates, alloy composition, and main phases of the matrix. In particular, the difference in the morphology of graphite precipitates causes the difference in the thermal conductivity between the ductile and gray cast irons. Gray cast irons are known to have an extensive 3-dimensional interconnected network of graphite having a larger thermal conductivity than that of the matrix, and the heat current can flow in the network. In contrast, ductile cast irons contain isolated graphite nodules in the matrix, and the heat current flows mainly in the matrix.[48] This fact explains our result that the thermal conductivity of the gray cast iron is larger than that of the ductile cast iron. On the other hand, $\Pi_{AEE}$ is determined by the electronic transport in the magnetic metal matrix because graphite precipitates are non-magnetic. To clarify the origin of the difference in $\Pi_{AEE}$ between the ductile and gray cast irons, it is necessary to perform systematic measurements of magneto-transport properties using various grades of cast irons with different compositions and microstructures.

Finally, we discuss strategies to further enhance $Z_{AEE}T$ in the iron-carbon alloys. In the case of the cast irons, the large $\Pi_{AEE}$ value dominantly contributes to the enhancement of $Z_{AEE}T$. However, the corresponding $S_{ANE}$ value of 0.8 μVK$^{-1}$ in the ductile cast iron is still smaller than those in other iron-based alloys.[17,18] The systematic investigation on the carbon-concentration dependence of AEE/ANE in the iron-carbon alloys is thus necessary. According to Refs. 17 and 18, $S_{ANE}$ in the FeAl (FeGa) alloys gradually increases with increasing the Al (Ga) composition and takes maximum value of 3.4 μVK$^{-1}$ (2.4 μVK$^{-1}$) at the Al (Ga) concentration of 19 at.% (32 at.%). In contrast, the carbon concentration in the cast irons is only < 8 at.%, and developing iron-carbon alloys with the carbon concentration of > 8 at.% may be a promising approach to improve the anomalous Ettingshausen/Nernst coefficient and the figure of merit. However, this is impossible in bulk alloys because of the solubility limit in the iron-carbon equilibrium phase diagram, but may be possible in thin films. As demonstrated in Ref. 49, non-equilibrium alloy films beyond solubility limits can be formed by a combinatorial sputtering method, and the high-throughput screening method established here is useful for determining the best composition in non-equilibrium iron-carbon alloys for AEE/ANE. On the other hand, reducing $\kappa$ through the suppression of the nonelectronic thermal conduction carried by phonons and/or magnons[50] is also an important approach for increasing the figure of merit.[51-55] The inset to Fig. 3(d) shows the nonelectronic contribution in the thermal conductivity $\Delta\kappa = \kappa - \kappa_{el}$ in the iron-carbon alloys and pure iron, where $\kappa_{el}$ is the electronic thermal conductivity estimated based on the Wiedemann-Franz law with the free-electron Lorenz number of 2.44 × 10$^{-8}$ (WΩK$^{-2}$).

Although the ductile cast iron shows the maximum value of $Z_{AEE}T$, its $\Delta\kappa$ is two times larger than that in the pure iron, indicating the potential of decreasing $\Delta\kappa$ by phonon and/or magnon engineering. Importantly, the cast irons have microstructures consisting of graphite defects, as discussed above. By designing and fabricating their microstructures to scatter phonons and magnons, $\kappa$ can be reduced by the suppression of $\Delta\kappa$. Our result suggests that $Z_{AEE}T$ can further be improved by the optimization of the composition and microstructures in the iron-carbon alloys.

In summary, we investigated the anomalous Ettingshausen effect (AEE) in iron-carbon alloys at room temperature. The absolute value of the anomalous Ettingshausen coefficient in the cast irons is an order of magnitude larger than that in the steel and pure iron. We found that the dimensionless figure of merit for AEE in the ductile cast iron is $1.3 \times 10^{-5}$, which is 55 times larger than that in the pure iron, owing to the increase of the anomalous Ettingshausen coefficient. Our result suggests a potential of the iron-carbon alloys as transverse magneto-thermoelectric materials, although the optimizations of the composition and microstructures are still necessary.

The authors thank K. Takamori, R. Iguchi, and Y. Kamada for valuable discussions and M. Isomura and K. Suzuki for technical supports. This work was supported by CREST "Creation of Innovative Core Technologies for Nano-enabled Thermal Management" (JPMJCR17I1) from JST, Japan; Grant-in-Aid for Research Activity start-up (20K22476) and Grant-in-Aid for Early-Career Scientists (21K14519) from JSPS, Japan; and NIMS Joint Research Hub Program. R.M. is supported by JSPS through the "JSPS Postdoctoral Fellowship for Research in Japan (Standard)" (P21064).

## AUTHOR DECLARATIONS
**Conflict of Interest**

The authors have no conflicts to disclose.

## DATA AVAILABILITY

The data that support the findings of this study are available from the corresponding author upon reasonable request.

|  | Compositions (at.%) | | | | | | |
|---|---|---|---|---|---|---|---|
|  | Fe | C | Si | Mn | P | S | Mg |
| Ductile cast iron | 81.075 | 13.598 | 4.935 | 0.229 | 0.047 | 0.014 | 0.102 |
| Gray cast iron | 81.849 | 13.232 | 4.582 | 0.248 | 0.066 | 0.023 | - |
| Steel | 98.466 | 0.647 | 0.257 | 0.586 | 0.036 | 0.009 | - |

**Table I.** The compositions of the iron-carbon alloys used in this study.

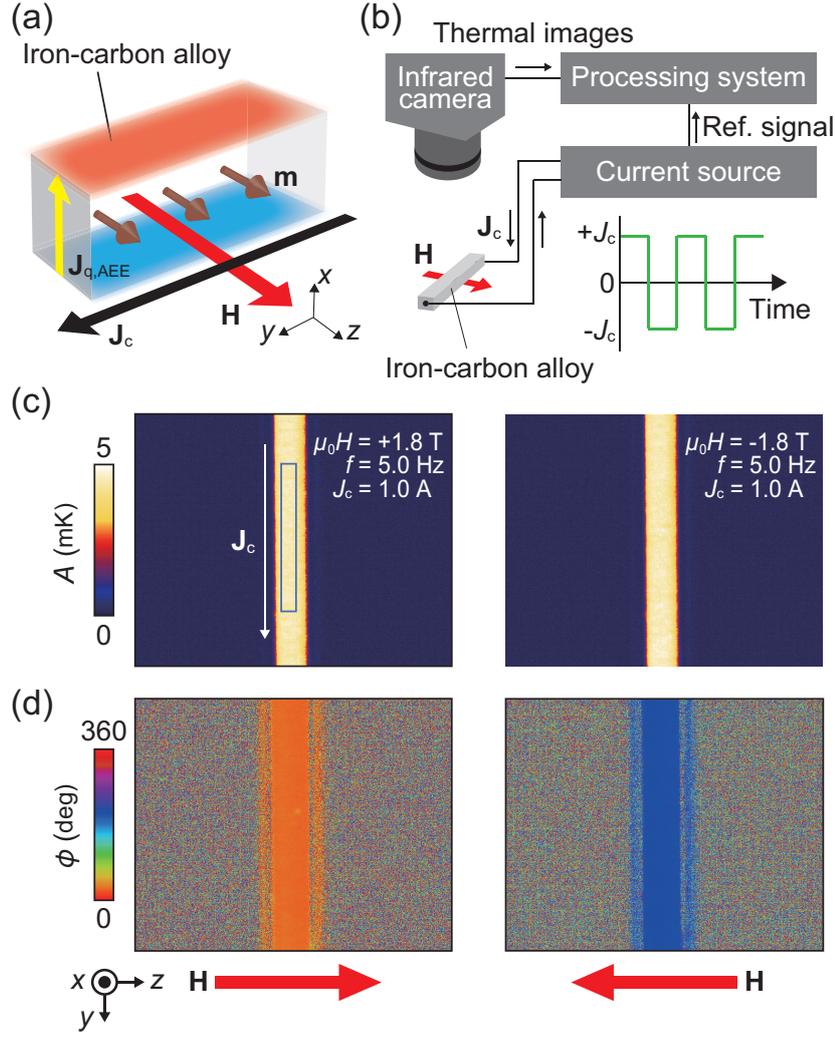

**FIG. 1.** (a) A schematic illustration of AEE in an iron-carbon alloy. $\mathbf{J}_c$, $\mathbf{J}_{q,\text{AEE}}$, $\mathbf{H}$, and $\mathbf{m}$ denote the charge current applied to the sample, heat current driven by AEE, direction of the applied magnetic field, and unit vector of the magnetization, respectively. (b) A schematic illustration of the setup for the LIT measurements. The lock-in amplitude $A$ (c) and phase $\phi$ (d) images for the ductile cast iron at $T = 301$ K, $f = 5.0$ Hz, and $J_c = 1.0$ A. $f$ and $J_c$ denote the frequency and amplitude of the square-wave-modulated charge current, respectively. The left and right images show the results measured at $\mu_0 H = +1.8$ T and $-1.8$ T, respectively. $\mu_0$ is the vacuum permeability. The results in Figs. 2 and 3 are obtained by averaging the temperature modulation signals on the area defined by the blue rectangle with the size of 30 × 300 pixels in (c). The temperature of the sample surface was determined from steady-state thermal images, which were simultaneously obtained with LIT images.

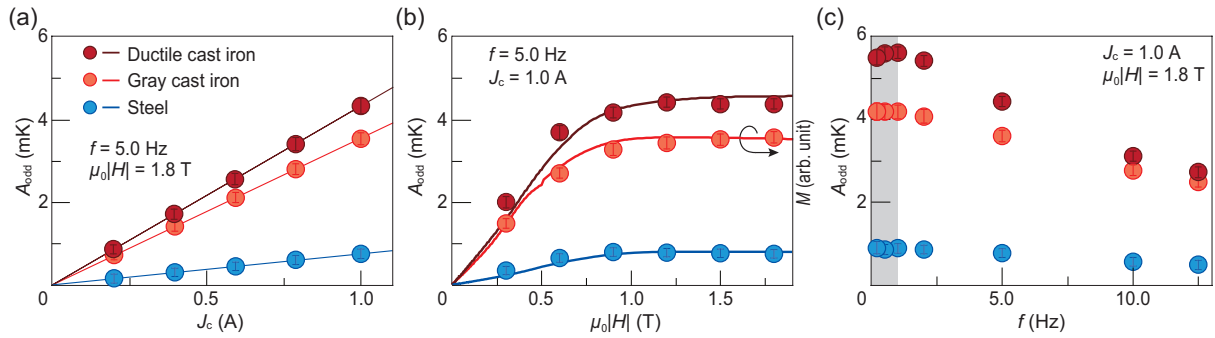

**FIG. 2.** $J_c$ (a), $\mu_0|H|$ (b), and $f$ (c) dependences of the $A_{odd}$ signals for the ductile cast iron (deep red data points), gray cast iron (light red data points), and steel (blue data points). $A_{odd}$ is the $H$-odd component of the lock-in amplitude. Solid lines in (a) show the results of linear fitting. In (b), the magnetization $M$ curves are also shown as solid curves. The gray shading in (c) represents the $f$ regime used for estimating the steady-state value of $A_{odd}$.

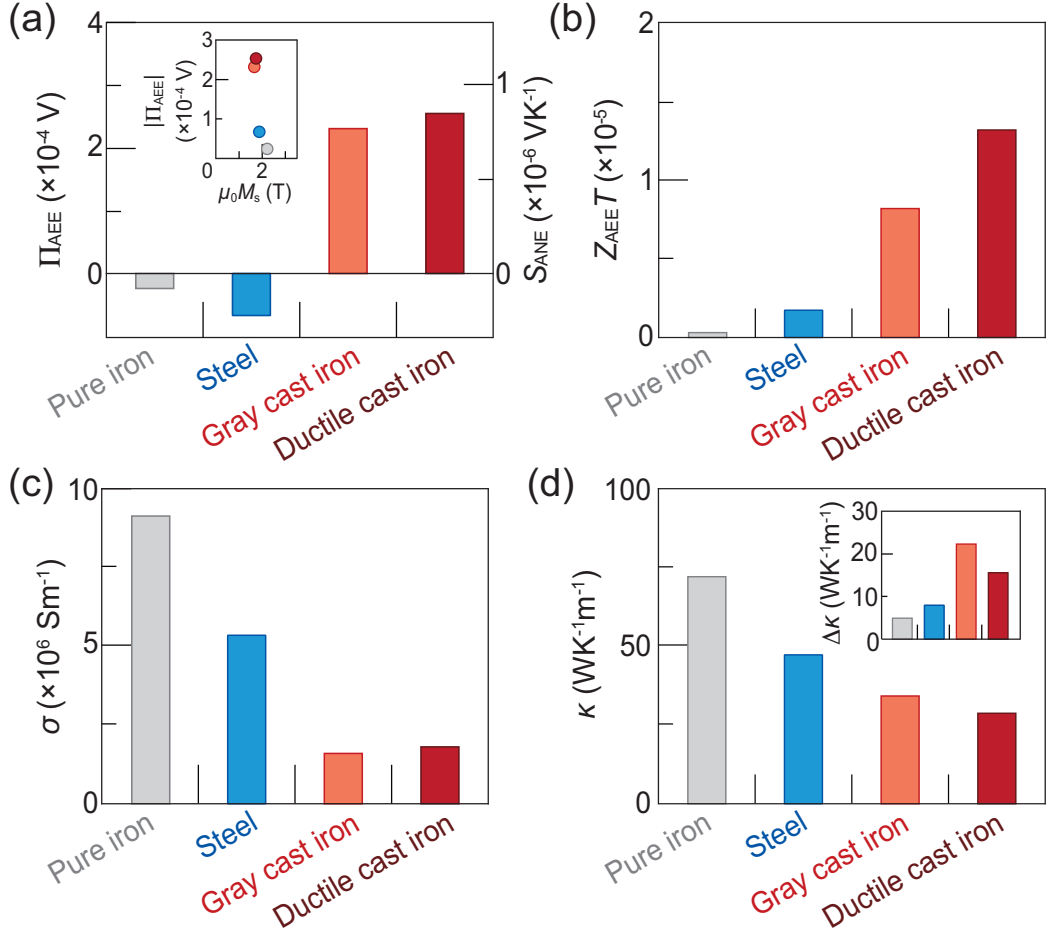

**FIG. 3.** The anomalous Ettingshausen coefficient $\Pi_{AEE}$ and anomalous Nernst coefficient $S_{ANE}$ (= $\Pi_{AEE}/T$) (a), dimensionless figure of merit $Z_{AEE}T$ (b), electrical conductivity $\sigma$ (c), and thermal conductivity $\kappa$ (d) for the pure iron (light gray) (Ref. 13), steel (blue), gray cast iron (light red), and ductile cast iron (deep red). The insets to (a) and (d) show the $\mu_0 M_s$ dependence of $|\Pi_{AEE}|$, where $M_s$ is the saturation magnetization, and $\Delta\kappa$ the nonelectronic thermal conductivity for the iron-carbon alloys and pure iron. The value of $\mu_0 M_s$ of the pure iron was taken from Ref. 6.

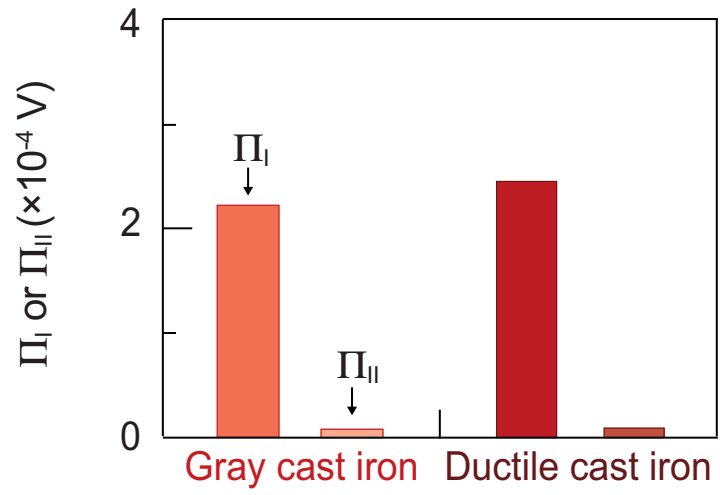

**FIG. 4.** $\Pi_\text{I}$ and $\Pi_\text{II}$ of the gray and ductile cast irons at $T$ = 301 K. $\Pi_\text{I}$ and $\Pi_\text{II}$ denote two different contributions in the anomalous Ettingshausen coefficient, defined in the main text.